\documentclass[aps,prb,amsmath,superscriptaddress,showpacs,amssymb,floatfix,twocolumn]{revtex4}
\usepackage{bbold}
\usepackage{graphicx}
\usepackage{subfigure}
\usepackage{color}
\usepackage{hyperref}

\begin{document}

\title{Axionic field theory of (3+1)-dimensional Weyl semi-metals}

\author{Pallab Goswami}
\affiliation{National High Magnetic Field Laboratory,
Florida State University, Tallahassee, FL 32306, USA}
\author{Sumanta Tewari}
\affiliation{Department of Physics and Astronomy, Clemson University, Clemson, SC
29634, USA}

\begin{abstract}
From a direct calculation of the anomalous Hall conductivity and an effective electromagnetic action obtained via Fujikawa's chiral rotation technique, we conclude that an axionic field theory with a non-quantized coefficient describes the electromagnetic response of the $(3+1)$-dimensional Weyl semi-metal. The coefficient is proportional to the momentum space separation of the Weyl nodes. Akin to the Chern-Simons field theory of quantum Hall effect, the axion field theory violates gauge invariance in the presence of the boundary, which is cured by the chiral anomaly of the surface states via the Callan-Harvey mechanism. This provides a unique solution for the radiatively induced CPT-odd term in the electromagnetic polarization tensor of the Lorentz violating spinor electrodynamics, where the source of the Lorentz violation is a constant axial four vector term for the Dirac fermion. A direct linear response calculation also establishes anomalous thermal Hall effect and a Wiedemann-Franz law, but thermal Hall conductivity does not directly follow from the well known formula for the gravitational chiral anomaly.

\end{abstract}


\maketitle

\vspace{10pt}

\section{Introduction} Usually topological states possess gapped spectrum in the bulk. The charge and the spin quantum Hall states in (2+1)-dimensions, and the strong topological insulator in (3+1) dimensions are some of the well known examples of such gapped topological states of matter. Topological considerations have also been applied to (3+1)-dimensional Weyl fermions, which describe a gapless, semi-metallic phase with nontrivial momentum space topology.\cite{Volovik1} In the simplest realization of Weyl semi-metal, there are linearly dispersing, 2-component massless Dirac fermions around a pair of diabolic points in the momentum space, described by the Hamiltonians $\hat{h}_{\pm}=\pm v_F \boldsymbol \sigma . (\mathbf{k}-\mathbf{K}_{\pm})$, where $v_F$ is the Fermi velocity, and $\boldsymbol \sigma$ is the vector formed out of three Pauli matrices.  The two component Hamiltonians $\hat{h}_{\pm}$ respectively describe the source (monopole) and the sink (anti-monopole) of the Berry curvature in the momentum space. The $\pm$ signs also correspond to the right and the left chirality of the Weyl fermions. For $\Delta \mathbf{k}=\mathbf{K}_{+}-\mathbf{K}_{-}\neq 0$, the right and the left Weyl fermions are separated in the momentum space and the semi-metal is topologically non-trivial. In contrast $\Delta \mathbf{k}=\mathbf{K}_{+}-\mathbf{K}_{-}= 0$ corresponds to topologically trivial, 4-component massless Dirac fermion. A Weyl semi-metal as described above, breaks time reversal symmetry and gives rise to anomalous charge and thermal Hall conductivities, and also possesses surface states with chiral dispersion in the directions orthogonal to $\Delta \mathbf{k}$. \cite{Vishwanath}

Recently the (3+1) dimensional Weyl semi-metal phase has been proposed for a variety of condensed matter systems, such as antiferromagnetically ordered pyrochlore iridates\cite{Vishwanath}, topological insulator-normal insulator multi-layers\cite{Burkov1, Burkov2, Burkov3}, magnetically doped topological insulator\cite{Cho,Qi} etc. In these systems electric charge is a conserved quantity, and the hallmark of topology can be found in anomalous electrodynamic properties\cite{Burkov1, Burkov2, Burkov3, Ran, Ninomiya, Aji, Son1, Son2}. For example (3+1)-dimensional chiral anomaly, which occurs for non-orthogonal arrangement of electric and magnetic fields may lead to a large negative magneto-resistance\cite{Ninomiya, Aji, Son1,Son2}. An anomalous charge Hall effect has also been predicted\cite{Burkov1, Burkov2, Burkov3, Ran} when the crystallographic symmetry does not cause any subtle cancelation of the Berry's phase among different pairs of Weyl fermions\cite{Ran}. On the other hand, Weyl quasiparticles can also occur for superfluids and superconductors\cite{Volovik1, Sau, Balents, Congjun}. For Weyl quasiparticles of superfluids and superconductors charge is not a conserved quantity. Therefore there will be no anomalous charge Hall effect. The topological property of Weyl superconductor will rather manifest through an anomalous thermal Hall effect\cite{Balents}. There is also tremendous current interest in high energy physics community in understanding how field theory anomalies affect the relativistic hydrodynamics and various transport quantities\cite{Surokawa,Kharzeev, Landsteiner, Loganayagam}.

Conventionally the low energy, universal physics of a gapped topologically ordered state is described by a topological field theory, which involves the conserved currents and the associated gauge fields. The topological field theories possess quantized coefficients, and describe the quantized non-dissipative response of the conserved charge. For example the charge quantum Hall effect in (2+1)-dimensions is described by a Chern-Simons term for the electromagnetic field\cite{Wilczek,Wen}
\begin{equation}
S[\mathcal{A}]=\frac{\sigma_{xy}}{4} \int d^{3}x \; \epsilon^{\mu \nu \lambda} \; \mathcal{A}_{\mu} \; \mathcal{F}_{\nu \lambda}
\label{eq:1}
\end{equation}
where $\sigma_{xy}$ is the quantized Hall conductivity and $\mathcal{A}_{\mu}$ and $\mathcal{F}_{\nu \lambda}$ are respectively the electromagnetic vector potentials  and the field strengths. By varying $S[\mathcal{A}]$ with respect to $\mathcal{A}_{1}$ and $\mathcal{A}_{2}$, we obtain the quantized antisymmetric tensor of the Hall conductivity, and a subsequent variation with respect to $A_{0}$ leads to $n=\sigma_{xy}B$, which in turn provides the thermodynamic Streda formula\cite{Streda} $\sigma_{xy}=\partial n/\partial B$, where $n$ is the density of the matter field. In the presence of a boundary, the Chern-Simons action becomes gauge non-invariant and the offending contribution due to the boundary is a (1+1)-dimensional term, which is proportional to the applied electric field along the boundary ($\propto \sigma_{xy}F_{02}/2\pi=\sigma_{xy}E_y/2\pi$, if the boundary is along y-direction). This gauge non-invariant contribution due to the boundary is precisely canceled by the chiral anomaly term of the gapless, chiral edge states via Callan-Harvey mechanism.\cite{Callan, Wen} This is how a topological field theory captures (i) the quantized response of the conserved charge in the bulk and (ii) the bulk-boundary correspondence.

For a (3+1)-dimensional, time reversal invariant, strong topological insulator the electromagnetic response is described by the axion term\cite{Zhang}
\begin{equation}
S[\mathcal{A}]=\frac{\theta \; e^2}{32 \pi^2} \int d^{4}x \; \epsilon^{\mu \nu \rho \lambda} \; \mathcal{F}_{\mu \nu} \; \mathcal{F}_{\rho \lambda}
\label{eq:2}
\end{equation}
where $\theta=0, \pi$ respectively describe the quantized magneto-electric coefficients of a topologically trivial and a non-trivial insulator. In the presence of a boundary between the topological and the trivial insulators, the axion angle $\theta$ discontinuously changes from $\pi$ to $0$, and gives rise to a gauge non-invariant boundary term. The boundary term in this case becomes a Chern-Simons term with $\sigma_{xy}=e^2/4\pi$, and is precisely canceled by the parity anomaly term\cite{Deser, Semenoff, Redlich} of the (2+1)-dimensional, two component Dirac fermions on the surface, via Callan-Harvey mechanism\cite{Callan, Fradkin1, Fradkin2, Zhang}. The above action describes a quantized magnetoelectric effect, which manifests as a half-integer quantum Hall effect of the two component Dirac fermions on the surface\cite{Zhang}.

Therefore it is natural to ask if there is a topological field theory description of a Weyl semi-metal. In the language of relativistic field theory, the momentum space separation between the left and right Weyl fermions is captured by a purely \emph{space-like} axial vector, and an axial vector breaks Lorentz invariance and CPT symmetry\cite{Volovik2}. The consequences of Lorentz violation due to CPT odd axial vector have been widely studied in high energy physics\cite{Colladay,Jackiw, Perez,Andrianov}. On the symmetry grounds it is well known that the spinor field can radiatively induce a CPT odd axion term in the vacuum polarization tensor\cite{Volovik2,Jackiw,Perez,Andrianov}. However the precise coefficient and the expression of the axion angle has been a subject of dispute.

Recently in Ref.~\onlinecite{Burkov4}, Zyuzin \& Burkov have applied Fujikawa's chiral rotation method\cite{Fujikawa} for deriving effective electromagnetic action of a Weyl semi-metal and have found a spatially varying axion angle $\theta= \Delta \mathbf{k}.\mathbf{x}$, where $\mathbf{x}$ describes the three spatial coordinates in the bulk. For an axial 4-vector $b_{\mu}$ and zero Dirac mass, the induced axionic action is given by
\begin{equation}
S_{W}^{ax}[\mathcal{A}]=\frac{e^2}{16 \pi^2}\int d^{4}x \: b_{\alpha}x^{\alpha} \: \epsilon^{\mu \nu \rho \lambda} \: \mathcal{F}_{\mu \nu}\; \mathcal{F}_{\rho \lambda}
\label{eq:ax}
\end{equation}
From this effective action they have reproduced the anomalous Hall conductivity $\sigma^{jk}=e^2/(2\pi^2)\epsilon^{ijk}b_{i}$, and the chiral magnetic conductivity $\sigma_{ch}=e^2/(2\pi^2)b_0$. However all the subtle aspects of chiral anomaly have not been addressed. In Ref.~\onlinecite{Grushin}, without dealing with the chiral anomaly, the bulk-boundary correspondence for anomalous Hall conductivity has been invoked to fix the ambiguity of the axion term for a general axial 4-vector in the presence of a Dirac mass. However these papers have not answered the most important question: why do the available calculations of vacuum polarization tensor\cite{Volovik2,Jackiw,Perez,Andrianov} fail to obtain the correct coefficient of the axionic term?

In this paper we thoroughly investigate various aspects of chiral anomaly and the axionic electrodynamics of (3+1)-dimensional Weyl semi-metal. When the external electric and magnetic fields are orthogonal, there is no (3+1) dimensional chiral anomaly. In this case we show that the (1+1) dimensional chiral anomaly associated with the surface states precisely cancel the gauge symmetry violating boundary term arising from the axionic action via Callan-Harvey mechanism\cite{Callan}, and the net theory indeed remains \emph{anomaly free}. When the electric and the magnetic fields are not orthogonal or a magnetic field is applied in the presence of a chiral chemical potential (purely \emph{time-like} axial vector), (3+1)-dimensional chiral anomaly shows up and this bulk chiral anomaly is not compensated by any surface effects. We also deal with the direct calculation of the CPT odd part of the polarization tensor and the anomalous Hall effect. From our detailed calculations it will become clear that, due to the violation of both Lorentz and spatial rotational invariance by a purely \emph{space-like} axial vector, any Lorentz or spatially rotational symmetric ultraviolet regularization leads to erroneous answer for anomalous Hall coefficient (inconsistent with the bulk-boundary correspondence), and renders the overall theory \emph{anomalous}. Only when the ultraviolet regularization is in accordance with the reduced rotational symmetry, we obtain the correct anomalous Hall conductivity, and simultaneously satisfy the bulk-boundary correspondence.

Our paper is organized as follows. In Sec. II we describe the relevant Dirac Hamiltonian augmented by an axial vector, and discuss the emergent spectrum in detail. In Sec. III we provide the detailed solutions of the chiral surface states of a Weyl semi-metal and a Weyl superconductor. We also demonstrate how (1+1)-dimensional chiral anomaly becomes embedded in a (2+1)-dimensional set up. Based on the surface state solutions we provide the relevant formula for anomalous charge Hall conductivity. In Sec. IV we obtain the axion electrodynamics of Weyl semi-metal by employing Fujikawa's chiral rotation technique. We show how this axionic field theory captures all the important features of the anomalous electrodynamics (i) anomalous Hall effect, (ii) Streda formula and (iii) bulk-boundary correspondence. In Sec. V we compute the anomalous Hall conductivity using Kubo formula, and show the detailed comparison among the answers obtained with different ultra-violet regularizations. In this section we also study the dependence of the anomalous Hall conductivity on the chemical potential. In Sec VI we briefly address the problem of purely \emph{time-like} axial vector or an axial chemical potential, and associated formula for the chiral magnetic conductivity. In Sec. VII we briefly comment on the relation between thermal Hall conductivity and the gravitational chiral anomaly. In Sec. VIII we present a summary of our results and the future directions, and a solution for the chiral surface states of a Weyl superconductor is presented in the Appendix.

\section{Model Hamiltonian of Weyl semi-metal} For simplicity we consider the following real-time (Minkowski space) action involving one species of 4-component Dirac fermion, described by
\begin{eqnarray}
S_0=\int d^4x \bar{\Psi}\bigg[i\gamma^{\mu}\partial_{\mu}-m-b_{\mu}\gamma^{\mu} \gamma^{5}\bigg]\Psi.
\label{eq:4}
\end{eqnarray}
where $\hbar$ and the Fermi velocity $v$ have been set to unity, $m$ is the conventional Dirac mass (Lorentz scalar), and the anticommuting $\gamma$ matrices satisfy $\{\gamma^{\mu},\gamma^{\nu}\}=2 g^{\mu \nu}$, $g^{\mu \nu}=(1,-1,-1,-1)$ is the metric tensor, $\gamma^{5}=i\gamma^0\gamma^1\gamma^2\gamma^3$ and $\bar{\Psi}=\Psi^{\dagger}\gamma_0$. We have added a Lorentz symmetry violating axial 4-vector term $\bar{\psi} b_{\mu}\gamma^{\mu}\gamma^{5}\psi$ in the action for generality. The Hamiltonian corresponding to the action in Eq.~\ref{eq:4} is described by
\begin{equation}
\mathcal{H}=\int d^3x \Psi^{\dagger}\left[-i\gamma^0 \gamma^j \partial_j +m\gamma^0+b_{\mu}\gamma^0\gamma^{\mu}\gamma^5\right]\Psi
\label{eq:5}
\end{equation}
The complete solution for the dispersion relation in the presence of a general axial 4-vector and Dirac mass has to be obtained numerically. The exact analytical solution can be found for (i) a general axial 4-vector and $m=0$, (ii) a purely \emph{space-like} axial 3-vector and $m \neq 0$ (iii) a purely \emph{time-like} axial vector or an axial chemical potential and $m \neq 0$ and (iv) a \emph{light-like} axial 4-vector and $m \neq 0$. A purely \emph{space-like} axial vector term breaks time reversal symmetry ($\mathcal{T}$), but preserves inversion ($\mathcal{P}$) and charge conjugation ($\mathcal{C}$) symmetries, and consequently breaks $CPT$ symmetry. An axial chemical potential $b_0$ also breaks $CPT$ symmetry, but in a different way. The axial chemical potential breaks $\mathcal{P}$, but preserves $\mathcal{T}$ and $\mathcal{C}$. In this section and also in the following Secs. III-V we will be mainly interested in a purely \emph{space-like} axial vector, i.e, $b_0=0$, which is pertinent for the realization of Weyl semi-metal. The effects of axial chemical potential $b_0$ will be discussed in Sec VI.

In the basis of even and odd parity bands (also known as Dirac representation of $\gamma$ matrices),
\begin{equation}{\gamma^0} =
 \left(\begin{array}{c c}
\mathbb{1} & 0 \\
0 & - \mathbb{1}
\end{array}\right), \;
{\gamma^j} =
 \left(\begin{array}{c c}
0 & {\sigma^j} \\
{-\sigma^j} & 0
\end{array}\right), \; {\gamma^5} =
 \left(\begin{array}{c c}
0 & \mathbb{1} \\
\mathbb{1} & 0
\end{array}\right), \;
\label{eq:6}
 \end{equation}
the Hamiltonian operator in the momentum space becomes
\begin{equation}
\hat{H}_{D}=\begin{bmatrix} m \: \mathbb{1}+\boldsymbol \sigma .\mathbf{b} & \boldsymbol \sigma .\mathbf{k} \\ \boldsymbol \sigma .\mathbf{k} & -m \: \mathbb{1}+\boldsymbol \sigma .\mathbf{b} \end{bmatrix}.
\label{eq:7}
\end{equation}
Thus an axial 3-vector term in Dirac basis, corresponds to a ferromagnetic order or uniform Zeeman coupling, where both parity even and odd bands have identical $g$ factors. If we use the chiral representation of the $\gamma$ matrices,
\begin{equation}{\gamma^0} =
\left(\begin{array}{c c}
0 & \mathbb{1} \\
\mathbb{1} & 0
\end{array}\right) , \;
{\gamma^j} =
 \left(\begin{array}{c c}
0 & {\sigma^j} \\
{-\sigma^j} & 0
\end{array}\right), \; {\gamma^5} =
\left(\begin{array}{c c}
-\mathbb{1} & 0 \\
0 & \mathbb{1}
\end{array}\right) \;
\label{eq:8}
\end{equation}
the Hamiltonian operator in the momentum space becomes
\begin{equation}
\hat{H}_{ch}=\begin{bmatrix} \boldsymbol \sigma .(\mathbf{k}+\mathbf{b}) & m \: \mathbb{1} \\ m \: \mathbb{1} & -\boldsymbol \sigma .(\mathbf{k}-\mathbf{b}) \end{bmatrix}
\label{eq:9}
\end{equation}

When $m=0$, the above Hamiltonian operators describe two Weyl fermions separated in the momentum space by the vector $\Delta \mathbf{k}=2\mathbf{b}$, which becomes particularly transparent from Eq.~\ref{eq:9}. In this case the $\hat{H}_{ch}$ is block diagonal, and chirality of the Weyl fermions after momentum space splitting is still determined by $\gamma^5$. For $m \neq 0$, the eigenstates of $\gamma^5$ get mixed, and the dispersion relations are given by
\begin{equation}
E_{k,s=\pm1}=\pm \sqrt{\left(\mathbf{k}\times \hat{b}\right)^2 +\left(|\mathbf{b}|+ s \sqrt{m^2+(\mathbf{k}. \hat{b})^2}\right)^2}
\label{eq:10}
\end{equation}
Due to the time reversal symmetry breaking, the Kramers degeneracy of the conduction and the valence bands are removed, which is captured by the non-degeneracy of conduction and valence bands for a given value of $s$. The bands corresponding to $s=+1$ are always fully gapped. But the bands corresponding to $s=-1$ are fully gapped, only when $m^2>\mathbf{b}^2$, and the system is an insulator. For $m^2<\mathbf{b}^2$, the conduction and the valence bands corresponding to $s=-1$, cross at $\mathbf{k}=\pm \hat{b} \sqrt{\mathbf{b}^2-m^2}$, giving rise to right and left Weyl fermions, with $\Delta \mathbf{k}=2\mathbf{b}\sqrt{1-m^2/b^2}$. In the subspace of these emergent Weyl fermions, we can define a new chirality matrix $\tilde{\gamma}^5$, and the low energy Hamiltonian will acquire the form
\begin{equation}
\hat{H}^{reduced}_{ch}=\begin{bmatrix} \boldsymbol \sigma .\left(\mathbf{k}+\hat{b}\sqrt{b^2-m^2}\right) & 0  \\ 0  & -\boldsymbol \sigma .\left(\mathbf{k}-\hat{b}\sqrt{b^2-m^2}\right) \end{bmatrix}
\label{eq:11}
\end{equation}

\section{Chiral surface states and (1+1)-dimensional chiral anomaly} In this section we find explicit solutions for the surface states of a Weyl semi-metal. To the best of our knowledge the explicit solution for the surface states of the Hamiltonian in Eq.~\ref{eq:5} has not been published in the literature. Therefore we provide the detailed solution of the surface states. In addition we also demonstrate for the first time how (1+1)-dimensional chiral anomaly becomes embedded into a (2+1) dimensional setup, and provides a strong constraint on the bulk effective field theory.

Consider $\hat{H}_{ch}$, with $\mathbf{b}=b\hat{z}$ and a boundary in the $yz$ plane. For simplicity we can model the boundary by a Heaviside step function of coordinate $x$, $b(\mathbf{x})=b \: \theta(x)$. To obtain the surface state solution we solve the following eigenvalue equation
\begin{equation}
\left(\begin{array}{c c}
i \boldsymbol \sigma \cdot \nabla +\sigma_3 \: b \: \theta(x) & m \: \mathbb{1} \\
m \: \mathbb{1} & -i \boldsymbol \sigma \cdot \nabla+\sigma_3 \: b \: \theta(x)
\end{array}\right) \Psi(\mathbf{x})=E\Psi(\mathbf{x}) ,
\label{eq:12}
\end{equation}
The momentum along $\hat{y}$ and $\hat{z}$ directions are good quantum numbers and after we substitute $\Psi(\mathbf{x})= e^{ik_y y +ik_z z} \psi(x)$ in Eq.~\ref{eq:12}, we find that the spinor $\psi(x)$ satisfies
\begin{widetext}
\begin{equation}
\left(\begin{array}{c c}
i \: \sigma_1 \: \partial_x -\sigma_2 k_y -\sigma_3 k_z+\sigma_3 \: b \:\theta(x) & m \mathbb{1} \\
m \mathbb{1} & -i \: \sigma_1 \: \partial_x +\sigma_2 k_y +\sigma_3 k_z +\sigma_3\: b \: \theta(x)
\end{array}\right) \psi(x)=E\psi(x). \;
\end{equation}
\label{eq:13}
\end{widetext}
It turns out that only the state corresponding to eigenvalue $+1$ of the matrix $\gamma^0\gamma^2$ leads to a normalizable solution. The general ansatz for such a state is
\begin{equation}
\psi(x)=u(x) \begin{pmatrix} 0 \\ 0 \\ 1 \\ i \end{pmatrix}+ v(x) \begin{pmatrix} 1 \\ -i \\ 0 \\0  \end{pmatrix}
\label{eq:14}
\end{equation}
which is a linear combination of the two eigenstates of the chiral matrix $\gamma^5$.
The above ansatz for $\psi(x)$ leads to the dispersion $E=+k_y$ and the functions $u(x)$ and $v(x)$ satisfy the following coupled differential equations
\begin{eqnarray}
\left[\partial_x + k_z+ b \theta(x)\right]u(x)+m v(x)=0 \label{eq:15}\\
m u(x)+ \left[\partial_x - k_z+ b \theta(x)\right]v(x)=0 \label{eq:16}
\end{eqnarray}
From these equations by eliminating $v(x)$, we obtain the following second order differential equations for $u(x)$,
\begin{eqnarray}
&& \partial_{x}^2 u(x)-\left( k_{z}^{2}+m^2 \right)u(x)=0, \: \forall  \: x<0 \label{eq:17}\\
&& \partial_{x}^2 u(x)+2b \: \partial_x u(x)-\left( k_{z}^{2}+m^2-b^2 \right)u(x)=0, \: \forall \: x>0 \nonumber \label{eq:18}\\
\end{eqnarray}
From these equations we obtain,
\begin{eqnarray}
u(x<0)&=&A \: e^{\sqrt{k_{z}^{2}+m^2} \: x} \label{eq:19}\\
v(x<0)&=&-\frac{1}{m}\left(k_z+\sqrt{k_{z}^{2}+m^2}\right)A \: e^{\sqrt{k_{z}^{2}+m^2}\: x} \label{eq:20}\\
u(x>0)&=&B \: e^{\left(-b+\sqrt{k_{z}^{2}+m^2}\right)x}+C \: e^{-\left(b+\sqrt{k_{z}^{2}+m^2}\right)x}\label{eq:21}\\
v(x>0)&=&-\frac{1}{m}\left(k_z+\sqrt{k_{z}^{2}+m^2}\right)\bigg \{B\: e^{\left(-b+\sqrt{k_{z}^{2}+m^2}\right)x} \nonumber \\ && +C \: e^{-\left(b+\sqrt{k_{z}^{2}+m^2}\right)x}\bigg \}\label{eq:22}
\end{eqnarray}
The continuity conditions on $u(x)$ and $v(x)$ at $x=0$ lead to $A= B+ C$ and $A=B+\frac{k_z-\sqrt{k_{z}^{2}+m^2}}{k_z+\sqrt{k_{z}^{2}+m^2}}C$, from which we obtain $C=0$, and $A=B$. Therefore the final solutions for $u(x)$ and $v(x)$ are given by
\begin{eqnarray}
u(x<0)&=&A \: e^{\sqrt{k_{z}^{2}+m^2} \: x} \label{eq:23}\\
u(x>0)&=&A \: e^{\left(-b+\sqrt{k_{z}^{2}+m^2}\right)x} \label{eq:24}\\
v(x)&=&-\frac{A}{m}\left(k_z+\sqrt{k_{z}^{2}+m^2}\right) u(x) \label{eq:25}
\end{eqnarray}
For $x>0$, the solution is normalizable only when $-\sqrt{b^2-m^2}<k_z<\sqrt{b^2-m^2}$. These surface state solutions with $E=+k_y$ are responsible for producing Fermi arc in the ARPES measurements \cite{Vishwanath}.

If we modify the Hamiltonian with a particle-hole symmetry breaking higher gradient term $\Psi^{\dagger} \partial_{j}^{2} \Psi$, the energy $E$ can  can acquire explicit dependence on $k_z$ \cite{Burkov2}, but chiral dispersion can occur only along $\hat{x}$ (in $xz$ plane) and $\hat{y}$ (in $yz$ plane) directions. Such perturbations are generically present in a lattice realization of Weyl semi-metal, and their qualitative effects in the low energy and momentum limit will be to modify $E=+v_F k_y$ to $E \approx \frac{k_{z}^{2}}{2m^*}+v_Fk_y$, where we have restored the Fermi velocity of the bulk-quasiparticles, and $m^*$ is a non-universal effective mass. In the presence of an electric field along $\hat{y}$ direction, the chiral dispersion along $\hat{y}$ leads to $U(1)$ \emph{chiral anomaly}. For any allowed value of $k_z$, we have chiral anomaly of one dimensional states $e^2/2\pi E_{x/y}$, and the net contribution from the entire range of $-\sqrt{b^2-m^2}<k_z<\sqrt{b^2-m^2}$ is given by $e^2 E_{x/y}/2\pi \int dk_z/ (2\pi) =e^2/(2\pi^2)\sqrt{b^2-m^2}E_{x/y}$. Thus (1+1)-dimensional chiral anomaly is embedded in a (2+1)-dimensional setup, and we can express this as the following effective action
\begin{equation}
S_{an}^{surface}=\frac{e^2}{2\pi^2}\sqrt{b^2-m^2}\int dt \: dy \: dz  f(y,z,t) E_y(y,z,t)
\label{eq:26}
\end{equation}
where the function $f(y,z,t)$ is related to a gauge transformation performed on the vector potential (which is also a chiral gauge transformation in this case). Under the variation with respect to $f$, we will obtain the anomalous Ward identity for the charge current (which is also the chiral current in this case). The presence of the chiral surface states and the associated U(1) chiral anomaly suggests that the bulk electromagnetic action contains some kind of topological term, which in the presence of a boundary can cancel the chiral anomaly to make the overall theory anomaly free\cite{Callan}. Akin to the edge modes of the quantum Hall problem, the chiral surface states give rise to the following anomalous Hall conductivity\cite{Burkov1,Burkov2}
\begin{equation}
\sigma_{xy}=\frac{e^2}{2\pi^2}\sqrt{b^2-m^2}
\end{equation}

\section{Axion electrodynamics in the bulk via chiral rotation} In this section we describe the bulk electromagnetic action obtained via Fujikawa's chiral rotation technique in Ref.~\onlinecite{Burkov4}. For simplicity we choose $m=0$, and two Weyl fermions are now located at $\mathbf{k}=\pm \mathbf{b}$. The relevant fermion action coupled to electromagnetic gauge fields is given by
\begin{eqnarray}
S[\Psi,\bar{\Psi},b,\mathcal{A}]=\int d^4x \bar{\Psi}\left[i\gamma^{\mu}(\partial_{\mu}+i\mathcal{A}_{\mu})-b_j\gamma^j\gamma^5\right]\Psi \nonumber \\
\label{eq:fg}
\end{eqnarray}
We can perform a chiral rotation of the spinor fields by an angle $\theta=\mathbf{b}.\mathbf{x}$, which removes the axial vector bilinear from the fermion sector in the action $S[\Psi,\bar{\Psi},b,\mathcal{A}]$ in Eq.~\ref{eq:fg}. Under the chiral rotation the spinor fields transform to $\Psi \to \exp (i \mathbf{b}.\mathbf{x}\gamma^5) \Psi^{'}$ and $\bar{\Psi} \to \bar{\Psi}^{'} \exp (i \mathbf{b}.\mathbf{x}\gamma^5)$. However, due to the non-invariance of the path integral measure under such a chiral rotation, the effective action acquires a contribution from the Jacobian of the transformation. This procedure leads to the following effective action
\begin{eqnarray}
S[\Psi,\bar{\Psi},b,\mathcal{A}]=\int d^4x \bar{\Psi}^{'}i\gamma^{\mu}(\partial_{\mu}+i\mathcal{A}_{\mu})\Psi^{'} \nonumber \\ +\frac{e^2}{16 \pi^2}\int d^4x \ \mathbf{b}.\mathbf{x} \ \epsilon^{\mu \nu \rho \lambda} \ \mathcal{F}_{\mu \nu} \ \mathcal{F}_{\rho \lambda}
\end{eqnarray}
The primed spinor fields represent topologically trivial 4-component massless Dirac fermions, and the entire effects of the axial vector is now captured by the axionic term for the electromagnetic fields,
\begin{equation}
\Delta S= \frac{e^2}{16 \pi^2}\int d^4x \ \mathbf{b}.\mathbf{x} \ \epsilon^{\mu \nu \rho \lambda} \ \mathcal{F}_{\mu \nu} \ \mathcal{F}_{\rho \lambda}
\end{equation}
This term does not reflect chiral anomaly in three-dimensional bulk. Rather this term is analogous to the condensation energy derived for a broken symmetry phase. It is more useful to write the gauge covariant form of this action
\begin{equation}
\Delta S=\frac{e^2 b_{\mu}}{4 \pi^2}\left(1-\delta_{\mu , 0}\right)\int d^4x \ \epsilon^{\mu \nu \rho \lambda} \ \mathcal{A}_{\nu} \ \partial_{\rho} \mathcal{A}_{ \lambda}
\label{eq:31}
\end{equation}
obtained via integration by parts. The variation of $\Delta S$ with respect to the vector potential provides the difference between the currents in the topological and the trivial sectors. Since topologically trivial 4-component massless Dirac fermion does not have any anomalous transport property, the variation of $\Delta S$ uniquely specifies the anomalous transport properties of the Weyl fermions. For $\mathbf{b}=b\hat{z}$, by varying the above action with respect to $\mathcal{A}_1$ and $\mathcal{A}_2$, we obtain the Hall conductivity $\sigma_{xy}=\frac{e^2 b}{2\pi^2}$, and a subsequent variation with respect to $\mathcal{A}_0$ leads to the Streda formula\cite{Streda} $\sigma_{xy}=\partial n/\partial B$. In the absence of a boundary the axionic action is gauge invariant. However in the presence of a boundary (say $\mathbf{b}(\mathbf{x})=b\theta(x) \hat{z}$), this action is not invariant under the gauge transformation $\mathcal{A}_{\mu}(x) \to \mathcal{A}_{\mu}(x)-\partial_{\mu}f(x)$, and there is a boundary term
\begin{equation}
\Delta S_{boundary}=-\frac{e^2 b} {2\pi^2} \int d^4x \: \delta(x_1) \: f(x) \: \epsilon^{31\rho \lambda} \: \partial_{\rho} \mathcal{A}_{\lambda}
\end{equation}
which is precisely canceled by the chiral anomaly from the surface states presented in Eq.~\ref{eq:26}. For $m \neq 0$, we can integrate out the higher lying bands corresponding to $s=+1$ described in Eq.~\ref{eq:10}, and use an effective action corresponding to the Hamiltonian operator shown in Eq.~\ref{eq:11}. Then our results for $m=0$ problem, can be taken over by replacing $b \to \sqrt{b^2-m^2}$.

In high energy physics literature\cite{Volovik2,Jackiw,Andrianov} it has been argued that a detailed knowledge of the high energy sector (beyond QED) or a physical input is required to fix the coefficient of the axionic term. However, in the present case, notice that the (1+1)-dimensional chiral anomaly of the surface states does not depend on the precise ultra-violet form of the Hamiltonian. Rather the emergence of the chiral surface states is precisely connected to the low energy physics of the bulk Weyl fermions. Thus in the present case the condition of an overall anomaly free theory or the bulk-boundary correspondence, unambiguously fixes the coefficient of the axionic action for the case of a purely \emph{space-like} axial vector to be $b_{\mu}/(4\pi^2)\left(1-\delta_{\mu , 0}\right)$ (see Eq.~\ref{eq:31}). For the corresponding massive theory, the coefficient becomes $b_{\mu}/(4\pi^2)\left(1-\delta_{\mu , 0}\right)\sqrt{1-m^2/|\mathbf{b}|^2}\theta(|\mathbf{b}|^2-m^2)$.

\section{Anomalous Hall conductivity via Kubo formula and ultra-violet regularizations}
In this section we will calculate the anomalous Hall conductivity using Kubo formula, which requires a direct calculation of the CPT odd part of the vacuum polarization tensor. In order to succinctly describe all the subtleties involving ultraviolet regularization, we again choose the simplest case of $m=0$. The origin of the anomalous term in the electrodynamic action is rooted in the trace identity $Tr[\gamma^{\mu}\gamma^{\nu}\gamma^{\rho}\gamma^{\lambda}\gamma^5]=-4i\epsilon^{\mu \nu \rho \lambda}$. The real-time fermion propagator corresponding to the action in Eq.~\ref{eq:4}, for $m=0$ is given by
\begin{equation}
G(k)=\frac{i}{2}\bigg [ \frac{(k-b)_{\mu}\gamma^{\mu}}{(k-b)^2}(1-\gamma^5)+\frac{(k+b)_{\mu}\gamma^{\mu}}{(k+b)^2}(1+\gamma^5)\bigg]
\end{equation}
After substituting this propagator in the expression of the polarization tensor
\begin{eqnarray}
\Pi^{\mu \nu}(p)=-ie^2 \int \frac{d^4k}{(2\pi)^4} Tr[\gamma^{\mu}G(k)\gamma^{\nu}G(k+p)]
\end{eqnarray}
and evaluating the trace over the $\gamma$ matrices, we obtain the CPT-odd anomalous part of the polarization tensor
\begin{eqnarray}
\Pi^{\mu \nu}_{odd}(p)&=&2e^2\epsilon^{\mu \nu \rho \lambda} p_{\lambda} \int \frac{d^4 k}{(2\pi)^4} \bigg [ \frac{(k+b)_{\rho}}{(k+b)^2(k+b+p)^2}\nonumber \\
&&-\frac{(k-b)_{\rho}}{(k-b)^2(k-b+p)^2}\bigg]
\label{eq:poltensor}
\end{eqnarray}
The integral produces a term proportional to $b_{\rho}$. It is important to take the momentum cut-off along the direction of the axial vector to infinity, only at the end of the calculations. Otherwise a naive shift of $k_{\mu}$ will make this integral to vanish\cite{Jackiw,Perez}.

If we now set the external momentum $p$ in the integrand to be zero, and perform a Wick rotation to Euclidean space, and perform the $k$-integral in a Lorentz invariant way with a 4-momentum cutoff $\Lambda$, we obtain
\begin{eqnarray}
&&\lim_{p \to 0}\frac{\Pi^{\mu \nu}_{odd}(p)}{p_\lambda}=\frac{4e^2}{(2\pi)^4 b^2} \epsilon^{\mu \nu \rho \lambda}\: b_{\rho}\int_{0}^{\Lambda} k^3 \: dk \int_{0}^{\pi} \sin^{2}\theta_1 d\theta_1 \nonumber \\ && \int_{0}^{\pi}\sin \theta_2 d\theta_2 \int_{0}^{2\pi} d\theta_3 \: \frac{b^2+2k b \: \cos \theta_1}{(k^2+b^2+2kb \: \cos \theta_1)^2}\\
&&=\frac{2e^2}{(2\pi)^2 b^4} \epsilon^{\mu \nu \rho \lambda}\: b_{\rho} \int_{0}^{\Lambda} k^3 \: dk \: \theta(b-k) \nonumber \\
&&=\frac{e^2}{8\pi^2} \epsilon^{\mu \nu \rho \lambda}\: b_{\rho}
\end{eqnarray}
This gives rise to an anomalous Hall conductivity and a chiral magnetic conductivity, which are smaller than the correct answers by a factor of 4 (see below Eq.~\ref{eq:ax}).

For the calculation of Hall conductivity using Kubo formula, we will choose the finite temperature fermion propagator and also consider the finite density effects by incorporating a non-zero chemical potential $\mu$ in the propagator. We further choose a pure \emph{space-like} axial vector $b_{\mu}=(0,\mathbf{b})$, and the external Euclidean momentum $p_{\mu}=i \omega_n \delta_{\mu,0}$. After performing a Matsubara sum and an analytic continuation to the real frequency $i\omega_n \to \omega +i\delta$, the dynamic Hall conductivity is found to be
\begin{eqnarray}
&&\sigma^{ij}=e^2 \epsilon^{ijl}\int \frac{d^3k}{(2\pi)^3}\bigg[\frac{(\mathbf{k}+\mathbf{b})_l}{|\mathbf{k}+\mathbf{b}|\left(4(\mathbf{k}+\mathbf{b})^2-\omega^2\right)} \times \nonumber \\
&& \left \{\tanh \frac{\beta}{2}\left(|\mathbf{k}+\mathbf{b}|-\mu \right)+\tanh \frac{\beta}{2}\left(|\mathbf{k}+\mathbf{b}|+\mu \right)\right \} \nonumber \\ && +(\mathbf{b} \to -\mathbf{b})\bigg]
\end{eqnarray}
Since we are concerned with a non-dissipative conductivity, we obtain the same answer for the zero temperature dc Hall conductivity for both limits $\omega /T \to 0$ and $\omega /T \to \infty$. The zero temperature dc Hall conductivity for $\mu=0$ is given by
\begin{eqnarray}
\sigma^{ij}&=&\frac{e^2}{2}\epsilon^{ijl}\int \frac{d^3k}{(2\pi)^3}\left[\frac{(\mathbf{k}+\mathbf{b})_l}{|\mathbf{k}+\mathbf{b}|^{3/2}}-(\mathbf{b} \to -\mathbf{b})\right]
\end{eqnarray}
If we perform the 3-momentum integral in a rotationally symmetric way, we obtain as in Ref.~\onlinecite{Volovik2}
\begin{eqnarray}
\sigma^{ij}&=&\frac{e^2}{\mathbf{b}^2(2\pi)^2}\epsilon^{ijl}\mathbf{b}_l \int_{0}^{\Lambda} k^2 \: dk \int_{0}^{\pi} \sin \theta \: d\theta \times \nonumber \\
& & \frac{(\mathbf{b}^2+ k|\mathbf{b}| \cos \theta)}{(k^2+\mathbf{b}^2+2 k|\mathbf{b}| \cos \theta)^{3/2}}  \\
&=& \frac{2e^2}{|\mathbf{b}|^3(2\pi)^2}\epsilon^{ijl}\mathbf{b}_l \int_{0}^{\Lambda} k^2 \: dk \theta(|\mathbf{b}|-k) \nonumber \\
&=& \frac{e^2}{6\pi^2}\epsilon^{ijl}\mathbf{b}_l,
\end{eqnarray}
which is three times smaller than the correct answer (see below Eq.~\ref{eq:ax}). We notice that both the Lorentz invariance and the spatial rotational invariance are broken by $\mathbf{b}$. Therefore, the above two methods of regularization do not conform to underlying symmetry and lead to incorrect answers. Below we use a regularization that is consistent with the broken symmetries in the presence of $\mathbf{b}$, and obtain the correct result for anomalous Hall conductivity using the Kubo formula.

There is a rotational symmetry only in the plane perpendicular to $\mathbf{b}$, and if we perform the integral over $k$ in a cylindrically symmetric way, it will be consistent with the underlying symmetry. We shall choose a  cut-off $\Lambda$ along the $\hat{b}$ direction, and keep the cut-off in the perpendicular plane to be $\infty$. With this regularization we obtain
\begin{eqnarray}
\sigma^{ij}&=&\frac{e^2}{(2\pi)^3}\epsilon^{ijl}\hat{b}_l \int_{0}^{\infty} k_{\perp} dk_{\perp} \int_{-\Lambda}^{\Lambda} dk_{\parallel} \int_{0}^{2\pi} d\phi \times \nonumber \\
& & \frac{k_{\parallel}+|\mathbf{b}|}{\left[k_{\perp}^{2}+(k_{\parallel}+|\mathbf{b}|)^2\right]^{3/2}} \\
&=& \frac{e^2}{(2\pi)^2}\epsilon^{ijl}\hat{b}_l \int_{-\Lambda}^{\Lambda} dk_{\parallel} \: \mathrm{sgn}(k_{\parallel}+|\mathbf{b}|) \nonumber \\
&=&\frac{e^2}{2\pi^2}\epsilon^{ijl}\mathbf{b}_l \label{halleqn}
\end{eqnarray}
which is indeed the correct answer. Therefore we need to choose (i) a method of integration which is consistent with the reduced rotational symmetry; (ii) a finite cut-off along the axial vector, which can be sent to infinity at the end of calculations; and (iii) we can keep the cut-off's in the directions perpendicular to the axial vector to be infinity without any trouble. In the presence of a non-zero Dirac mass, the expressions become more involved. However, this physically motivated regularization does yield the correct anomalous Hall conductivity $\sigma^{ij}=e^2/(2\pi^2)\sqrt{\mathbf{b}^2-m^2}\epsilon^{ijl}\hat{b}_l$ for $\mathbf{b}^2>m^2$. For $\mathbf{b}^2<m^2$ the anomalous Hall conductivity vanishes, which is consistent with the absence of the chiral surface states, and certainly is the correct answer for the bulk insulator.

Now we consider the effects of finite density $(\mu \neq 0)$ on the anomalous Hall conductivity. For concreteness we choose $\mu>0$. The zero temperature dc Hall conductivity is then given by
\begin{eqnarray}
&&\sigma^{ij}=\frac{e^2}{(2\pi)^3}\epsilon^{ijl}\hat{b}_l \int_{0}^{\infty} k_{\perp} dk_{\perp} \int_{-\Lambda}^{\Lambda} dk_{\parallel} \int_{0}^{2\pi} d\phi \times \nonumber \\
&& \frac{k_{\parallel}+|\mathbf{b}|}{\left[k_{\perp}^{2}+(k_{\parallel}+|\mathbf{b}|)^2\right]^{3/2}} \: \theta (k_{\perp}^{2}+(k_{\parallel}+|\mathbf{b}|)^2-\mu^2)\\
&&=\frac{e^2}{(2\pi)^2}\epsilon^{ijl}\hat{b}_l \int_{-\Lambda+|\mathbf{b}|}^{\Lambda+|\mathbf{b}|} x \: dx \: \left[ \frac{\theta(|x|-\mu)}{|x|}+ \frac{\theta(\mu-|x|)}{\mu}\right] \nonumber \\
&&=\frac{e^2}{2\pi^2}\epsilon^{ijl}\hat{b}_l
\end{eqnarray}
This is an interesting result which implies that the anomalous Hall conductivity remains unchanged at a finite density. This is in accordance with the general principle that U(1) chiral anomaly remains unaffected by chemical potential or finite temperature \cite{Dolan, Gomez}. But, this result is only valid for unbounded linear dispersion of Dirac fermions. When we consider a more realistic dispersion of the fermions, such that the linear dispersion is smoothly cut-off by higher gradient terms, the contributions from the partially filled or occupied states generically remain finite. Therefore, the anomalous Hall conductivity of real materials will be modified by a non-universal contribution from the partially filled or empty states.

\begin{figure}[htb]
\includegraphics[width=8.5cm,height=6cm]{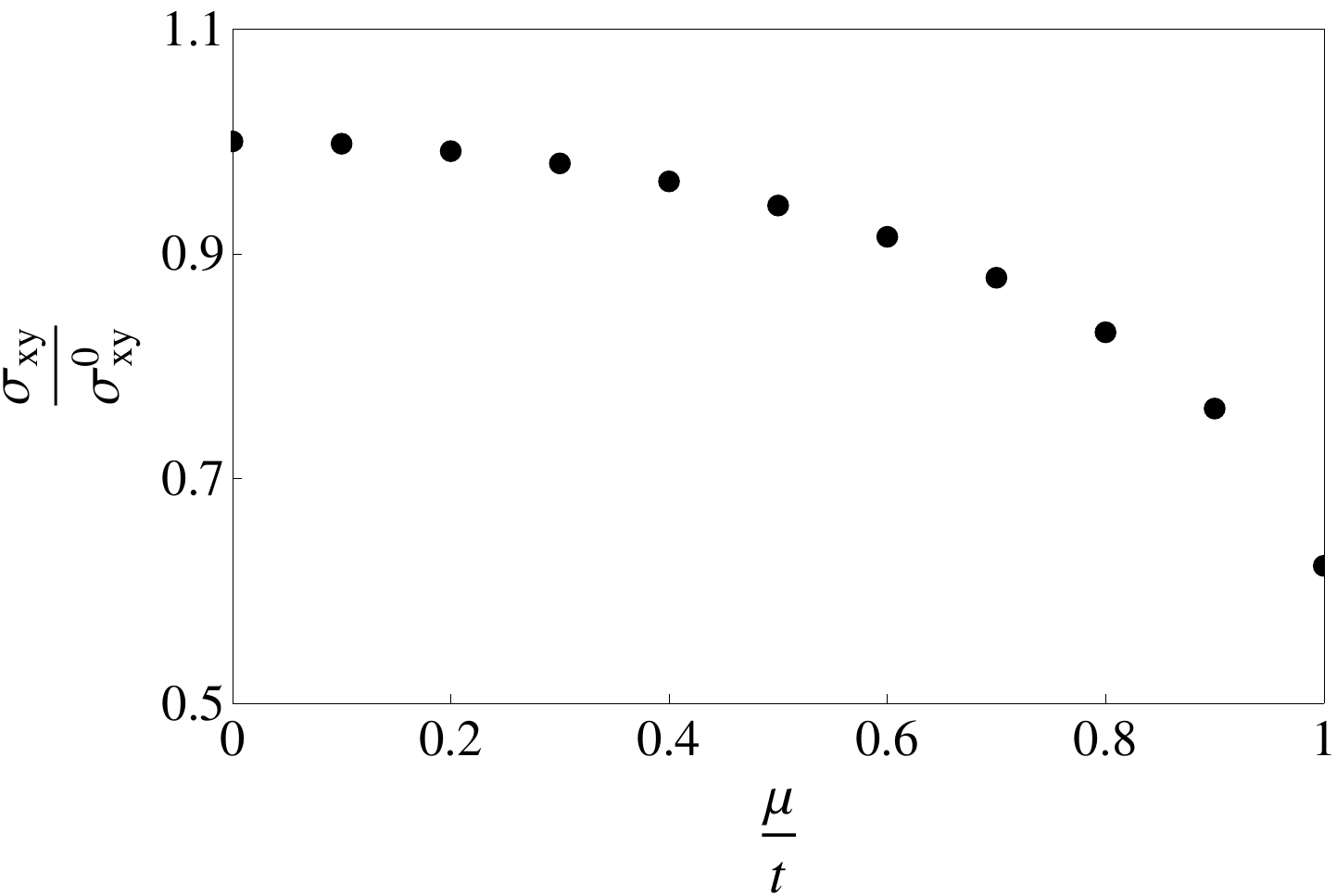}
\caption{(Color online) $\sigma_{xy}/\sigma^0_{xy}$ as a function of $\mu/t$, where $\sigma^0_{xy}=e^2/(4\pi)$ is the anomalous Hall conductivity at half-filling.}
\label{Hall}
\end{figure}

For justifying this conclusion we have used the following lattice model of Weyl fermions
\begin{eqnarray}
H= t(\sin k_1 \tau_1+ \sin k_2 \tau_2 + \cos k_3 \tau_3)+m(2-\cos k_1 \nonumber \\
 -\cos k_2)\tau_3=\mathbf{N}_{\mathbf{k}}\cdot \boldsymbol \tau,
\end{eqnarray}
at finite density. When $m >t/2$, the model supports only one pair of Weyl fermions. The right and the left handed Weyl points are respectively located at $(0,0, \pi/2)$ and $(0,0,-\pi/2)$, leading to $\mathbf{b}=(0,0,\pi/2)$. For $\mu=0$, the anomalous Hall conductivity is given by $\sigma^0_{xy}=e^2/(4\pi)$. The Hall conductivity at arbitrary density is given by the well known formula\cite{Haldane,Xiao}
\begin{eqnarray}\label{AHE}
\sigma_{xy}=e^2 \; \epsilon_{abc} \; \sum_n \int \frac{d^3k}{(2\pi)^3} \Omega_{\mathbf{k},n,z} \; f(E_{\mathbf{k},n}),
\end{eqnarray}
where $f(E)$ is the Fermi function and the Berry curvatures for two bands are determined as
\begin{eqnarray}\label{Berry}
\Omega_{\mathbf{k},n,a}=(-1)^n\epsilon_{abc}\frac{\mathbf{N}_{\mathbf{k}} \cdot \bigg[ \frac{\partial \mathbf{N}_{\mathbf{k}}}{\partial k_b}\times \frac{\partial \mathbf{N}_{\mathbf{k}}}{\partial k_c}\bigg]}{4|\mathbf{N}_{\mathbf{k}}|^3}.
\end{eqnarray}
We have numerically computed $\sigma_{xy}$ as a function of $\mu$ at $T=0$, by employing Eq.~(\ref{AHE}), and the results are shown in Fig.~\ref{Hall}. We have plotted the dimensionless quantity $\sigma_{xy}/\sigma^0_{xy}$ as a function of the dimensionless variable $\mu/t$. As $\mu/t$ grows, $\sigma_{xy}/\sigma^0_{xy}$ gradually decreases from unity. These results demonstrate the inadequacy of the linearized continuum theory for calculating the anomalous Hall current at finite density.

\section{Chiral Magnetic conductivity in the presence of axial chemical potential} For an axial chemical potential $b_0 \neq0$, the number of left and right fermions will be different, and the (3+1)-dimensional bulk chiral anomaly can manifest in the presence of a magnetic field. When $b_0 \neq 0$ and $m=0$, left and right Weyl fermions respectively produce particle and hole like Fermi surfaces (which will be unstable against interaction effects, due to the perfect nesting of the Fermi surfaces) to maintain charge neutrality. For finite chiral chemical potential, theory is truly anomalous and the bulk (3+1)-dimensional chiral anomaly in the presence of an electromagnetic field can not be removed by any surface state. It is also important to note that for a purely \emph{time-like} axial vector and $m=0$, there is no surface state solution. By chiral rotation method we shall find an axion angle $\theta=b_0 \: t$, which leads to an equilibrium current along the direction of the applied magnetic field $j_i =e^2b_0 B_i/(2\pi^2)$, and the chiral magnetic conductivity defined by $\sigma_{ch}=j_i/B_i=e^2b_0 /(2\pi^2)$.

For a purely \emph{time-like} axial vector, we have spatial rotational symmetry (but no Lorentz symmetry). Our method of regularization for the direct calculation of the chiral magnetic conductivity will be to choose: (i) a frequency cut-off $\Lambda$, (ii) keep the spatial cut-off to be infinity, and (iii) perform the spatial momentum integral in a rotationally symmetric manner. This method again produces the correct answer for the chiral magnetic conductivity $e^2b_0 /(2\pi^2)$. In the presence of a non-zero Dirac mass and an axial chemical potential, the dispersion relation is given by
\begin{equation}
E_{\mathbf{k},s=\pm1}=\pm \sqrt{(|\mathbf{k}|+ s \: b_0)^2+m^2},
\end{equation}
which describes an insulating bulk. Due to the broken inversion symmetry the Kramers degeneracy is absent. With our regularization, the chiral magnetic conductivity for a massive Dirac fermion in the presence of an axial chemical potential vanishes. Therefore we find, the chiral magnetic conductivity is finite for massless Dirac fermion in the presence of an axial chemical potential, while for the corresponding massive theory it vanishes. In Refs.~\onlinecite{Kharzeev, Andrianov, Burkov4}, the chiral magnetic conductivity of the massive problem has been found to be $e^2b_0 /(2\pi^2)$. However in these works, the frequency cut-off has been taken to be infinity and the spatial cut-off has been kept finite. Hence their regularization procedure is opposite to the one we have used.

\section{Thermal Hall conductivity and gravitational anomaly} Since an axial 3-vector breaks time reversal symmetry, on symmetry ground we expect an anomalous thermal Hall conductivity. A linear response calculation in the bulk (after accounting for diathermal contributions) leads to an Wiedeman-Franz law between anomalous charge and thermal Hall conductivities\cite{Niu1, Niu2} for a Weyl semi-metal. The thermal Hall conductivity $\kappa^{ij}$ is found to be
\begin{equation}
\kappa^{ij}= \frac{\pi^2 k_{B}^{2} T}{3 e^2}\sigma^{ij}= \frac{k_{B}^{2}T }{12} \epsilon^{ijl} \Delta \mathbf{k}_l,
\end{equation}
This result is also consistent with the heat transported by the chiral surface states\cite{Fisher,Cappelli}. For a Weyl superconductor there is no charge Hall conductivity. The anomalous thermal Hall conductivity for the ferromagnetic superconductor of described in Ref.~\onlinecite{Sau} is given by
\begin{equation}
\kappa_{sc}^{ij}= \frac{k_{B}^{2}T}{24}\epsilon^{ijl}\sum_{s}\Delta \mathbf{k}_{s,l},
\end{equation}
where $\Delta \mathbf{k}_{s}$ is the momentum space separation of the Weyl fermions corresponding to the spin projection $s$. In contrast to the Weyl semi-metal there is an additional pre-factor of $1/2$, which reflects that we are dealing with BdG quasiparticles in the bulk\cite{Read}. The pre-factor of $1/2$ is also consistent with the fact that surface states are now described by chiral Majorana fermions. Recently the gravitational chiral anomaly has been argued to be responsible for a quantized thermal Hall effect on the surface of a topological superconductor in class DIII\cite{Ryu,ZhangQi, RyuNagaosa}. However we find that the linear response thermal Hall conductivity is not directly captured by the well known gravitational anomaly formula, even though the gravitational anomaly formula correctly captures the bulk-boundary correspondence at zero temperature, and the anomaly formula also verifies the correct conformal charge of the surface states\cite{Goswami1}. Similar concerns regarding the connection between the thermal Hall effect and the gravitational chiral anomaly have also been raised in Refs.~\onlinecite{Kimura, Stone}.

\section{Summary and future directions} In this paper we have explored the relationship between the chiral anomaly and the effective electromagnetic action of a (3+1)-dimensional Weyl semi-metal in detail. We have obtained explicit solutions for the chiral surface states and associated (1+1)-dimensional chiral anomaly. Based on the chiral anomaly of the surface states we have demanded the existence of an anomalous bulk action, such that the overall theory becomes anomaly free via Callan-Harvey mechanism\cite{Callan}. From a subsequent calculation of the bulk charge Hall conductivity via Fujikawa's chiral rotation method and also by Kubo formula, we have established that the anomalous electrodynamic properties in the bulk is described by axion electrodynamics. Unlike the strong topological insulator, which has gapped spectrum and a quantized coefficient for axionic action, the coefficient of the axionic electrodynamics for Weyl semi-metal is proportional to the momentum space separation of the Weyl nodes. The momentum space separation of the Weyl fermions can be renormalized due to disorder and interaction effects. However, weak disorder and long range Coulomb interactions turn out to be irrelevant perturbations\cite{Goswami2,Hosur} for (3+1)-dimensional Dirac and Weyl fermions at zero chemical potential. Therefore at zero chemical potential, the renormalization effects on the anomalous Hall conductivity will become unimportant at low temperatures. However the combined effects of disorder and interaction on the anomalous Hall conductivity at a finite density is an interesting open problem.

We have pointed out the subtleties involved in the direct calculation of anomalous Hall conductivity by using Kubo formula, and also provided a regularization scheme that respects the underlying symmetry. Our regularization does produce the correct answer for the anomalous Hall conductivity for a Lorentz violating spinor electrodynamics, where the Lorentz violation is caused by a purely \emph{space-like} axial vector bilinear for the Dirac fermions. Our regularization provides the correct chiral magnetic conductivity for a massless Dirac fermion in the presence of an axial chemical potential or a purely \emph{time-like} axial vector. Based on our regularization we have found zero chiral magnetic conductivity for a massive Dirac fermion in the presence of an axial chemical potential, which is at odds with the result reported in the literature\cite{Kharzeev, Andrianov, Burkov4}. Recently, some authors have claimed that the chiral magnetic effect is an artefact of the linearized continuum theory and can not be realized in real materials\cite{Franz,Basar,Landsteinerch}. These claims have been subsequently disproved in Refs.~\onlinecite{Burkov5,Goswamitewari}, where finite chiral magnetic conductivities have been calculated based on different lattice realizations of Weyl semi-metal with an axial chemical potential.

A linear response calculation of the thermal Hall conductivity has shown the existence of a Wiedemann-Franz law between the charge and the thermal Hall conductivities. In contrast to the recent claims in the literature, we find that the thermal conductivity does not directly follow from the well known gravitational anomaly formula. The detailed connection between the thermal Hall effect and gravitational anomaly will be discussed in a separate publication\cite{Goswami1}.

We note that a non-quantized axion angle has also been found for the magneto-electric response of the doped topological insulators, when the chemical potential lies in the conduction or the valence bands \cite{Bergman,Barkeshli}. In contrast to the Weyl semi-metal, the non-quantized theta term of the doped topological insulators does not describe bulk anomalous Hall effect. Rather, it corresponds to a non-quantized anomalous surface Hall effect, when the time-reversal symmetry is broken on the surface by applying magnetic impurities. An interesting application of the Callan-Harvey effect has also been discussed in a recent work\cite{WangZhang}, when the Weyl semi-metal is destabilized through a charge density wave order.

\section{acknowledgements}  P. G. is supported at the National High Magnetic Field Laboratory by NSF Cooperative Agreement No.DMR-0654118, the State of Florida, and the U. S. Department of Energy. S. T. would like to thank DARPA MTO, Grant No. FA9550-10-1-0497 and NSF, Grant No. PHY-1104527 for support. We thank Kun Yang and Bitan Roy for discussions.

\appendix

\section{Surface states of Weyl superconductor} For a Weyl superconductor described in Ref.~\onlinecite{Sau}, we can use the Hamiltonian operators of Sec. II, as long as we identify $\Psi(\mathbf{x})$ to be a Nambu-spinor. However to make a direct connection we choose the ferromagnetic triplet superconductor described in Ref.~\onlinecite{Sau}, which has the same pairing symmetry as $He^3-A_2$ phase. Due to the ferromagnetic moment there are spin-split Fermi surfaces in the normal state. In addition the same spin pairing is chosen with the symmetry $k_x-ik_y$, but with distinct amplitudes for two spin projections. The reduced BCS Hamiltonian is described by
\begin{eqnarray}
\mathcal{H}_{BCS}&=&\sum_{s=\pm, \mathbf{k}}\bigg[\left(\frac{\mathbf{k}^2}{2m}-\mu_s \right)c_{\mathbf{k},s}^{\dagger}c_{\mathbf{k},s}+\frac{\Delta_s}{k_{F,s}}(k_x-ik_y) \nonumber \\&& \times c_{\mathbf{k},s}^{\dagger}c_{-\mathbf{k},s}^{\dagger}+h.c. \bigg],
\end{eqnarray}
where $\mu_s$, $k_{F,s}=\sqrt{2m \mu_s}$, and $\Delta_s$ are respectively the chemical potential, the Fermi surface radius, and the pairing amplitude for spin projection $s$. For each spin projection $s$, we have a pair of Weyl fermions located at $\mathbf{k}_s=\pm k_{F,s} \hat{z}$. In order to obtain a qualitative description of the surface states we choose a spatially varying chemical potential $\mu_s(\mathbf{x})=\mu_s \tanh(x/\xi)$\cite{Read}. Again we retain $k_y$ and $k_z$ as good quantum numbers, and replace $k_x$ by $-i\partial_x$. For small energy and momentum, we can ignore $\partial_{x}^2$ and $k_{y}^{2}$ in comparison to $\partial_x$ and $k_y$, but retain the only $k_z$ dependence through the quadratic term $k_{z}^{2}/2m$ to obtain
\begin{eqnarray}
\bigg[\left(\frac{k_{z}^2}{2m}-\mu_s\tanh \frac{x}{\xi} \right)\tau_3&+&\frac{\Delta_s}{k_{F,s}}\left(-i\tau_1 \partial_x+\tau_y k_y \right)\bigg] \psi_{s}(x)\nonumber \\
&&=E_s \psi_s(x)
\end{eqnarray}
In order to obtain a normalizable solution we need to choose $\psi_s(x)$ to be an eigenstate of $\tau_y$ with eigenvalue +1, which leads to the ansatz
$\psi_{s}^{T}(x)=u_s(x)(1, \: -i)$, and the dispersion $E_s=-\Delta_s k_y/k_{F,s}$. The function $u_s(x)$ satisfies
\begin{equation}
\partial_x u_s(x)-\frac{k_{F,s}}{\Delta_s}\left(\frac{k_{z}^2}{2m}-\mu_s\tanh \frac{x}{\xi} \right)u_s(x)=0,
\end{equation}
and the solution is described by
\begin{equation}
u_s(x)=u_{0,s}\exp \left[ \frac{k_{F,s}}{\Delta_s}\frac{k_{z}^{2}}{2m} x\right]\left( \cosh \frac{x}{\xi}\right)^{-\mu_s \xi}\ \,
\end{equation}
which has the following asymptotic property
\begin{eqnarray}
\lim_{|x| \to \infty} u_s(x) \sim u_{0,s} \exp \left[ \frac{k_{F,s}}{\Delta_s}\left(\frac{k_{z}^{2}}{2m}x-\mu_s |x|\right)\right]
\end{eqnarray}
The above solution $u_s(x)$ is normalizable only when $-k_{F,s}<k_z<k_{F,s}$, and describes a Majorana fermion with chiral dispersion along $-\hat{y}$ direction.

\end{document}